\documentclass[a4paper,12pt, showpacs, showkeys]{revtex4-1}
\usepackage{amsbsy}
\usepackage{amsmath}
\usepackage{amsfonts}
\usepackage{amsthm}
\usepackage{graphicx}

\usepackage[latin1]{inputenc}

\begin{document}
\author{Vladimir~Klinshov}
\email{vladimir.klinshov@gmail.com}
\affiliation{Institute of Applied Physics of the Russian Academy of Sciences, 46 Ul'yanov Street, 603950, Nizhny Novgorod, Russia} 
\affiliation{University of Nizhny Novgorod, 23 Prospekt Gagarina, 603950, Nizhny Novgorod, Russia}
\author{Leonhard~L{\"u}cken}
\email{leonhard.luecken@wias-berlin.de}
\affiliation{Weierstrass Institute, Mohrenstrasse 39, 10117 Berlin, Germany}
\author{Dmitry~Shchapin}
\affiliation{Institute of Applied Physics of the Russian Academy of Sciences, 46 Ul'yanov Street, 603950, Nizhny Novgorod, Russia} 
\affiliation{University of Nizhny Novgorod, 23 Prospekt Gagarina, 603950, Nizhny Novgorod, Russia}
\author{Vladimir~Nekorkin}
\affiliation{Institute of Applied Physics of the Russian Academy of Sciences, 46 Ul'yanov Street, 603950, Nizhny Novgorod, Russia} 
\affiliation{University of Nizhny Novgorod, 23 Prospekt Gagarina, 603950, Nizhny Novgorod, Russia}
\author{Serhiy~Yanchuk}
\affiliation{Weierstrass Institute, Mohrenstrasse 39, 10117 Berlin, Germany}

\title{Multistable jittering in oscillators with pulsatile delayed feedback}


\begin{abstract} 
Oscillatory systems with time-delayed pulsatile feedback appear in
various applied and theoretical research areas, and received a growing
interest in recent years. For such systems, we report a remarkable
scenario of destabilization of a periodic regular spiking regime.
At the bifurcation point numerous regimes with non-equal interspike
intervals emerge. We show that the number of the emerging, so-called
``jittering'' regimes grows \emph{exponentially} with the delay
value. Although this appears as highly degenerate from a dynamical
systems viewpoint, the ``multi-jitter'' bifurcation occurs robustly
in a large class of systems. We observe it not only in a paradigmatic
phase-reduced model, but also in a simulated Hodgkin-Huxley neuron
model and in an experiment with an electronic circuit.
\end{abstract}

\maketitle

Interaction via pulse-like signals is important in neuron populations
\cite{Intro1,Zillmer2007,Canavier2010}, biological \cite{Intro2,Winfree2001},
optical and optoelectronic systems \cite{Intro3}. Often, time delays
are inevitable in such systems as a consequence of the finite speed
of pulse propagation \cite{Intro4}. In this letter we demonstrate
that the pulsatile and delayed nature of interactions may lead to
novel and unusual phenomena in a large class of systems. In particular,
we explore oscillatory systems with pulsatile delayed feedback which
exhibit periodic regular spiking (RS). We show that this RS regime
may destabilize via a scenario in which a variety of higher-periodical
regimes with non-equal interspike intervals (ISIs) emerge simultaneously.
The number of the emergent, so-called ``jittering'' regimes grows
\emph{exponentially} as the delay increases. Therefore we adopt the
term ``multi-jitter'' bifurcation.

Usually, the simultaneous emergence of many different regimes is a
sign of degeneracy and it is expected to occur generically only when
additional symmetries are present \cite{Zillmer2007,Intro5}. However,
for the class of systems treated here no such symmetry is apparent.
Nevertheless, the phenomenon can be reliably observed when just a
single parameter, for example the delay, is varied. This means that
the observed bifurcation has codimension one \cite{Kuznetsov1995}.
In addition to the theoretical analysis of a simple paradigmatic model,
we provide numerical evidence for the occurrence of the multi-jitter
bifurcation in a realistic neuronal model, as well as an experimental
confirmation in an electronic circuit.

\begin{figure}[t]
\centering
\includegraphics[width=0.75\textwidth]{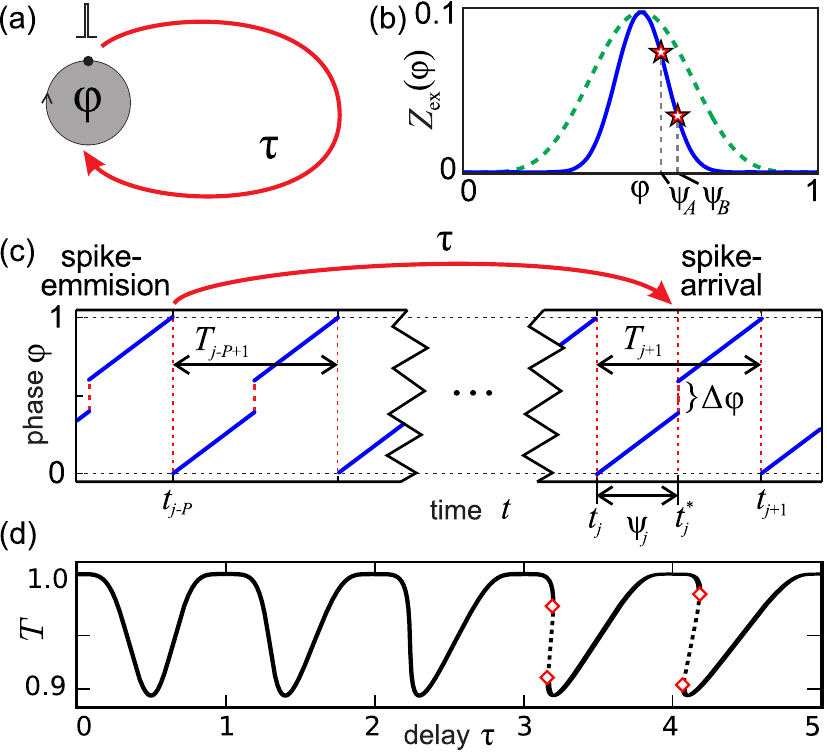}

\protect\caption{\label{fig:fig1}(a) Scheme of the model. (b) Shape
of the PRC $Z_{{\rm ex}}(\varphi)$ for $q=5$ (dashed green) and
$q=28$ (solid blue). The stars indicate points with slope $-1$.
(c) Construction of the map (\ref{eq:T_map}): the new ISI $T_{j+1}$
depends on the pulse emitted at $t=t_{j-P}$. (d) Periods $T$ of
the RS versus the delay $\tau$ for $q=5$, obtained from Eq.~(\ref{eq:T_RS}).
Solid lines indicate stable RS regimes, dashed unstable. Diamonds
indicate saddle-node bifurcations.}
\end{figure}
As a universal and simplest oscillatory spiking model in the absence
of the feedback, we consider the phase oscillator $d\varphi/dt=\omega$,
where $\varphi\in\mathbb{R}\left(\mod1\right)$, and $\omega=1$ without
loss of generality. When the oscillator reaches $\varphi=1$ at some
moment $t$, the phase is reset to zero and the oscillator produces
a pulse signal. If this signal is sent into a delayed feedback loop
{[}Fig.~\ref{fig:fig1}(a){]} the emitted pulses affect the oscillator
after a delay $\tau$ at the time instant $t^{*}=t+\tau$. When the
pulse is received, the phase of the oscillator undergoes an instantaneous
shift by an amount $\Delta\varphi=Z(\varphi(t^{\ast}-0))$, where
$Z(\varphi)$ is the phase resetting curve (PRC). Thus, the dynamics
of the oscillator can be described by the following equation \cite{Canavier2010,Goel2002,Klinshov2011,Lucken2012a,Luecken2013}:

\begin{eqnarray}
\frac{d\varphi}{dt} & = & 1+Z(\varphi)\sum\limits _{t_{j}}\delta(t-t_{j}-\tau),\label{eq:1}
\end{eqnarray}
where $t_{j}$ are the instants when the pulses are emitted. Note
that we adopt the convention that positive values of the PRC lead
to shorter ISIs. For numerical illustrations we use $Z_{{\rm ex}}(\varphi):=0.1\sin^{q}\left(\pi\varphi\right),$
where $q$ controls the steepness of $Z_{{\rm ex}}\left(\varphi\right)$
{[}see Fig.~\ref{fig:fig1}(b){]}. However, our analysis is valid
for an arbitrary amplitude or shape of the PRC.

In \cite{Klinshov2013} it was proven that a system with pulsatile
delayed coupling can be reduced to a finite-dimensional map under
quite general conditions. To construct the map for system (\ref{eq:1})
let us calculate the ISI $T_{j+1}:=t_{j+1}-t_{j}$. It is easy to
see that $T_{j+1}=1-Z(\psi_{j})$, where $\psi_{j}=\varphi(t_{j}^{\ast}-0)=t_{j}^{*}-t_{j}$
is the phase at the moment of the pulse arrival $t_{j}^{*}=t_{j-P}+\tau$
{[}Fig.~\ref{fig:fig1}(c){]}. Here, $P$ is the number of ISIs between
the emission time and the arrival time. Substituting $t_{j}=t_{j-P}+T_{j-P+1}+...+T_{j}$,
we obtain the ISI map
\begin{equation}
T_{j+1}=1-Z\left(\tau-\sum_{k=j-P+1}^{j}T_{k}\right).\label{eq:T_map}
\end{equation}

The most basic regime possible in this system is the regular spiking
(RS) when the oscillator emits pulses periodically with $T_{j}=T$
for all $j$. Such a regime corresponds to a fixed point of the map
(\ref{eq:T_map}) and therefore all possible periods $T$ are given
as solutions to
\begin{equation}
T=1-Z\left(\tau-PT\right),\label{eq:T_RS}
\end{equation}
where $P=\left[\tau/T\right]$, and hence $\tau-PT=\tau({\rm mod}T)$.
Figure~\ref{fig:fig1}(d) shows the period $T$ as a function of
$\tau$ for $Z_{{\rm ex}}(\varphi)$ and $q=5$.

To analyze the stability of the RS regime, we introduce small perturbations
$\delta_{j}$ such that $T_{j}=T+\delta_{j}$, and study whether they
are damped or amplified with time. The linearization of (\ref{eq:T_map})
in $\delta_{j}$ is straightforward and leads to the characteristic
equation
\begin{eqnarray}
\lambda^{P}-\alpha\lambda^{P-1}-\alpha\lambda^{P-2}-...-\alpha\lambda-\alpha & = & 0,\label{eq:char}
\end{eqnarray}
where $\alpha:=Z^{\prime}(\psi)$ is the slope of the PRC at the phase
$\psi=\tau\mod T$ (cf. \cite{Canavier2010,Foss2000}).

There are two possibilities for the multipliers $\lambda$ to become
critical, i.e. $\left|\lambda\right|=1$. The first scenario takes
place at $\alpha=1/P$ when the multiplier $\lambda=1$ appears, which
indicates a saddle-node bifurcation {[}diamonds in Fig.~\ref{fig:fig1}(d){]}.
In general these folds of the RS-branch lead to the appearance of
multistability and hysteresis between different RS regimes \cite{Foss2000,Foss1996,Hashemi2012}.
\begin{figure*}
\includegraphics[width=1\textwidth]{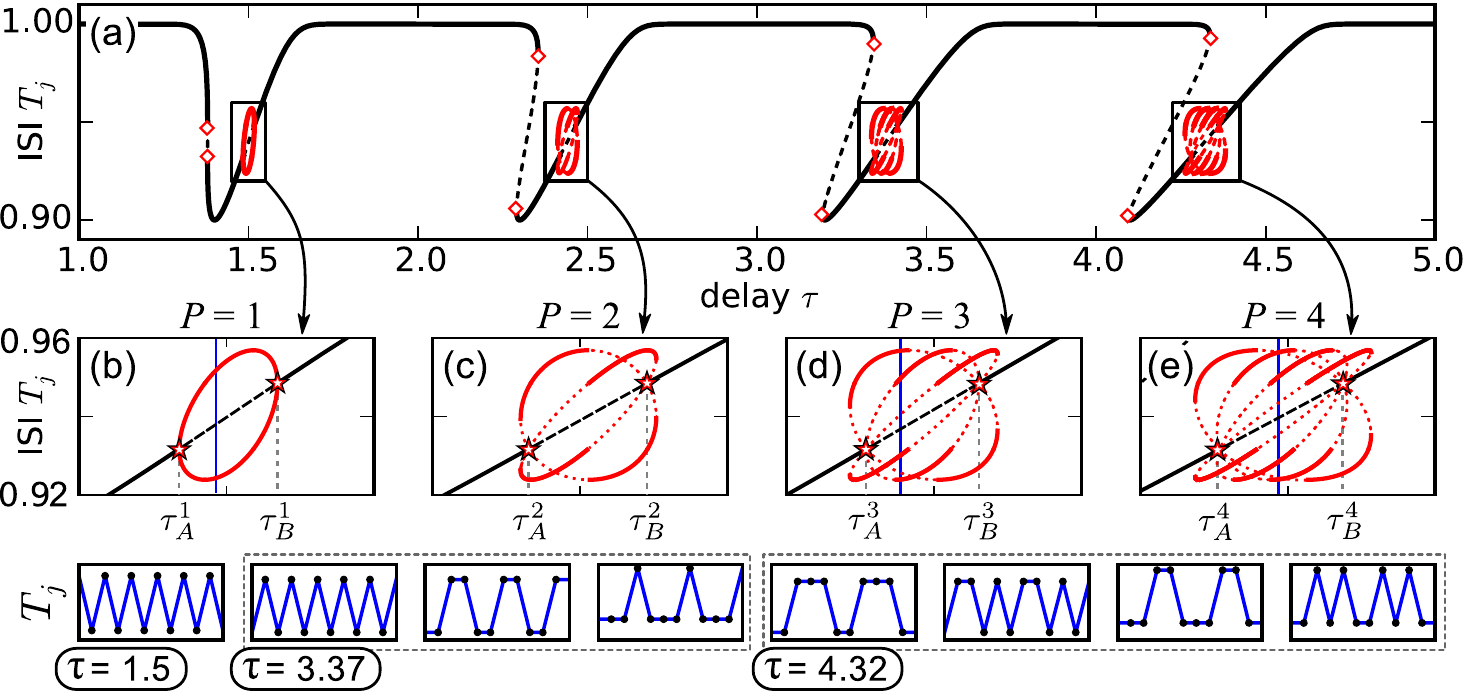}
\protect\caption{\label{fig:fig2}(a) ISIs $T_{j}$ versus the
delay $\tau$ for (\ref{eq:1}) with $Z=Z_{{\rm ex}}$ and $q=28$.
Solid lines correspond to stable, dashed and dotted to unstable solutions.
Black color indicates RS regimes, red stands for bipartite solutions
(with two different ISIs). Diamonds indicate saddle-node bifurcations
and stars multi-jitter bifurcations. (b), (c), (d), and (e) are zooms
of (a) for $P=1,2,3,4$. The lower panels show examples of stable
bipartite solutions, for which the corresponding sequences of ISIs
are plotted versus time. Solutions within the same dashed frame coexist
at a common value of $\tau$, which is also indicated by a vertical
blue line in the corresponding zoom.}
\end{figure*}

The second scenario is much more remarkable and takes place at $\alpha=-1$,
where $P$ critical multipliers $\lambda_{k}=e^{i2\pi k/(P+1)},$
$1\le k\le P$, appear simultaneously. This feature is quite unusual
since in general bifurcations one would not expect more than one real
or two complex-conjugate Floquet multipliers become critical at once
\cite{Kuznetsov1995}. In the following we study this surprising
bifurcation in detail and explain why we call it ''multi-jitter''.

In order to observe the multi-jitter bifurcation, the PRC $Z(\varphi)$
must possess points with sufficiently steep negative slope $Z^{\prime}(\varphi)=-1$.
For instance, in the case $Z(\varphi)=Z_{{\rm ex}}(\varphi)$, such
points exist for $q>q^{*}\approx27$. For such $q$, two points $\psi_{A},\psi_{B}\in\left(0,1\right)$
exist where $Z_{{\rm ex}}^{\prime}\left(\psi_{A,B}\right)=-1$ {[}see
stars in Fig.~\ref{fig:fig1}(b){]}. This means that for appropriate
values of the delay time $\tau$, such that $\tau\left({\rm mod}T\right)\in\{\psi_{A},\psi_{B}\}$,
it holds $\alpha=-1$, and the multi-jitter bifurcation takes place.
Using Eq.~(\ref{eq:T_RS}) one may determine the corresponding values
of $\tau$ for each possible $P\geq1$:
\begin{equation}
\tau_{A,B}^{P}=P(1-Z(\psi_{A,B}))+\psi_{A,B}.\label{eq:tau_ABP}
\end{equation}

Figure~\ref{fig:fig2} shows the numerically obtained bifurcation
diagram for $q=28$. All values of its ISIs $T_{j}$ observed after
a transient are plotted by solid lines versus the delay $\tau$. Black
lines correspond to RS regimes, while irregular regimes with distinct
ISIs are indicated in red color. Black dashed lines correspond to
unstable RS solutions obtained from Eq.~(\ref{eq:T_RS}). For the
intervals $\tau\in(\tau_{A}^{P},\tau_{B}^{P})$, the RS regime destabilizes
and several stable irregularly spiking regimes appear.

Let us study in more detail the bifurcation points $\tau=\tau_{A,B}^{P}$
for different values of $P$. For $P=1$, only one multiplier $\lambda=-1$
becomes critical. Note that in this case the map (\ref{eq:T_map})
is one-dimensional and the corresponding bifurcation is just a supercritical
period doubling giving birth to a stable period-2 solution existing
in the interval $\tau\in(\tau_{A}^{1},\tau_{B}^{1})$ {[}Fig.~\ref{fig:fig2}(b){]}.
For this solution the ISIs $T_{j}$ form a periodic sequence $(\overline{\Theta_{1},\Theta_{2}})$, 
where the periodicity of the sequence is indicated by an overline.
It satisfies
\begin{equation}
\Theta_{2}=1-Z\left(\tau-\Theta_{1}\right)\text{ and }\Theta_{1}=1-Z\left(\tau-\Theta_{2}\right).\label{eq:T1_T2}
\end{equation}
For $P\geq2$, $P$ multipliers become critical simultaneously at
$\tau=\tau_{A,B}^{P}$ and the RS solution is unstable for $\tau\in(\tau_{A}^{P},\tau_{B}^{P})$.
Numerical study shows that various irregular spiking regimes appear
in this interval. We observe solutions, which have ISI sequences
of period $P+1$ but exhibit only two different ISIs in varying order
{[}see Fig.~\ref{fig:fig2}, bottom{]}. As a result, each solution
corresponds to only two, and not $P+1$, points in Figs.~\ref{fig:fig2}(a),(c)--(e).
In the following we call such solutions ``bipartite''. For larger
$P$, a variety of different bipartite solutions with $(P+1)$-periodic
ISI sequences can be observed in $\tau\in(\tau_{A}^{P},\tau_{B}^{P})$.
The stability regions of these solutions alternate and may overlap
leading to multistable regimes  {[}see Appendix, Figs.\ref{fig:s1}--\ref{fig:s4}{]}. 

The bipartite structure of the observed solutions can be explained
by their peculiar combinatorial origin. Indeed, all bipartite solutions
can be constructed from the period-2 solution$(\overline{\Theta_{1},\Theta_{2}})$
existing for $P=1.$ Consider an arbitrary $(P+1)$-periodic sequence
of ISIs $(\overline{T_{1},T_{2},...,T_{P+1}})$, where each $T_{j}$
equals one of the solutions $\Theta_{1,2}$ of (\ref{eq:T1_T2}) for
some delay $\tau=\tau_{0}\in\left[\tau_{A}^{1},\tau_{B}^{1}\right]$.
Let $n_{1}\geq1$ and $n_{2}\geq1$ be the number of ISIs equal to
$\Theta_{1}$ and to $\Theta_{2}$ respectively. Then it is readily
checked that the constructed sequence is a solution of (\ref{eq:T_map})
at the feedback delay time
\begin{equation}
\tau_{n_{1},n_{2}}=\tau_{0}+\left(n_{1}-1\right)\Theta_{1}+\left(n_{2}-1\right)\Theta_{2}.\label{eq:tau-n1-n2}
\end{equation}
Red dotted lines in Figs.~\ref{fig:fig2}(c), (d), and (e) show the
branches of bipartite solutions constructed from (\ref{eq:T1_T2})
with $P=n_{1}+n_{2}-1=2,3,4$. Note that these solutions lie exactly
on the numerical branches which validates the above reasoning. However,
some parts of the branches are unstable and not observable. Since
each bipartite solution corresponds to a pair of points $\left(\tau_{n_{1},n_{2}},T_{1}\right)$
and $\left(\tau_{n_{1},n_{2}},T_{2}\right)$, solutions with identical
$n_{1}$ and $n_{2}$ correspond to the same points in the bifurcation
diagrams in Fig.~\ref{fig:fig2}. For instance, when $P=3$ the branches
corresponding to the solutions $(\overline{\Theta_{1},\Theta_{2},\Theta_{1},\Theta_{2}})=(\overline{\Theta_{1},\Theta_{2}})$
and $(\overline{\Theta_{1},\Theta_{1},\Theta_{2},\Theta_{2}})$ lie
on top of each other.

Let us estimate the number of different bipartite solutions for a
given $P\in\mathbb{N}$. The number of different binary sequences
of the length $P+1$ equals $2^{P+1}$. Subtracting the two trivial
sequences corresponding to the RS one gets $2^{P+1}-2$. Disregarding
the possible duplicates by periodic shifts (maximally $P+1$ per sequence)
one obtains an estimate for the total number ${\cal N}_P$ of bipartite solutions
for a given value of $P$ as
\begin{equation}
{\cal N}_P\ge\frac{2^{P+1}-2}{(P+1)}.\label{eq:nr-bipartite-slns}
\end{equation}
Notice that all these bipartite solutions exist at the same value
of $P$ but for different ranges of the delay $\tau$. Nevertheless,
all emerge from the RS solution in the bifurcation points $\tau_{A,B}^{P}$.
To see this, let us consider the limit $\tau_{0}\to\tau_{A}^{1}$.
In this case the ISIs $\Theta_{1}(\tau_{0})$ and $\Theta_{2}(\tau_{0})$
tend to the same limit $T_{A}=1-Z\left(\psi_{A}\right)$ which is
the period of the RS at the bifurcation point. Then, (\ref{eq:tau-n1-n2})
converges to $\tau_{A}^{1}+(n_{1}+n_{2}-2)T_{A}=\tau_{A}^{P}$, while
all bipartite solutions converge to the RS with period $T_{A}$. 

Thus, all bipartite solutions branch off the RS in the bifurcation
points $\tau_{A,B}^{P}$. This finding is clearly recognizable in
Figs.~\ref{fig:fig2}(c)--(e), where stars indicate the multi-jitter
bifurcation points. Numerical simulations show that many of the bipartite
solutions stabilize leading to high multistability. In particular,
we observe that all bipartite solutions with same values of $n_{1}$
and $n_{2}$ exhibit identical stability. This emergence of numerous
irregular spiking, or jittering, regimes motivates the choice of the
name ``multi-jitter bifurcation''.

High multistability is a well-known property of systems with time
delays. A common reason is the so-called reappearance of periodic
solutions \cite{Yanchuk2009}. This mechanism may cause multistability
of coexisting periodic solutions, whose number is linearly proportional
to the delay. Due to multi-jitter bifurcation, multistability can
develop much faster, since the number of coexisting solutions grows
exponentially with the delay {[}cf. Eq.~(\ref{eq:nr-bipartite-slns}){]}.
This suggests that the underlying mechanism is quite different. 

Irregular spiking regimes similar to the ones described here were
reported previously for systems exhibiting a dynamical ``memory effect'',
where the effect of each incoming pulse lasts for several periods
\cite{Glass}. In system (\ref{eq:1}), however, the effect of a
pulse decays completely within one period. Therefore the origin of
jittering must be different. In fact, it relies on another kind of
memory, which is provided by the delay line and stores the last $P$
ISIs. This memory preserves fundamental properties of time, which
are responsible for the degeneracy of the multi-jitter bifurcation.
To explain this let us consider (\ref{eq:T_map}) as a $P$-dimensional
mapping $\left(T_{j-P+1},...,T_{j}\right)\mapsto\left(T_{j-P+2},...,T_{j+1}\right).$
Disregarding the calculation of the new ISI $T_{j+1}$ as in (\ref{eq:T_map}),
all the map does is to move the timeframe by shifting all ISIs one
place ahead. The rigid nature of time allows no physically meaningful
modification of this part of the map which could unfold the degenerate
bifurcation. Moreover, the new ISI $T_{j+1}$ depends exclusively
on the sum of the previous ISIs which has the effect that all past
intervals have an equal influence on the new ISI regardless of their
order. As a consequence the combinatorial accumulation of coexisting
solutions with differently ordered ISIs is generated.

 Besides the delayed feedback, another essential ingredient for the
multi-jitter bifurcation is the existence of points where the PRC
fulfills $Z^{\prime}(\varphi)<-1$. Since the PRC is a characteristic
that can be measured for an arbitrary oscillator \cite{Winfree2001},
this condition gives a practical criterion for the occurence of jittering
regimes. In this context it is worth noticing that the condition $Z^{\prime}(\varphi)<-1$
is equivalent to the non-monotonicity of the system's response to
an external pulse, i.e. there are such phases $\varphi_{1}<\varphi_{2}$
that a pulse can reverse their order as $\varphi_{1}+Z(\varphi_{1})>\varphi_{2}+Z(\varphi_{2})$.
Note that for a smooth one-dimensional system as (\ref{eq:1}) the
reversal of phases is only possible if the feedback takes the form
of $\delta$-pulses. With pulses of finite duration two continuous
orbits connecting the different phase values before and after the
pulse cannot cross each other. This prevents a reversal of phases.
However, for oscillators with a phase space of dimension larger than
one the phase points $\varphi_{1}$ and $\varphi_{2}$ can exchange
their order without necessitating the orbits to intersect.

\begin{figure}[t]
\centering
\includegraphics[width=1.\textwidth]{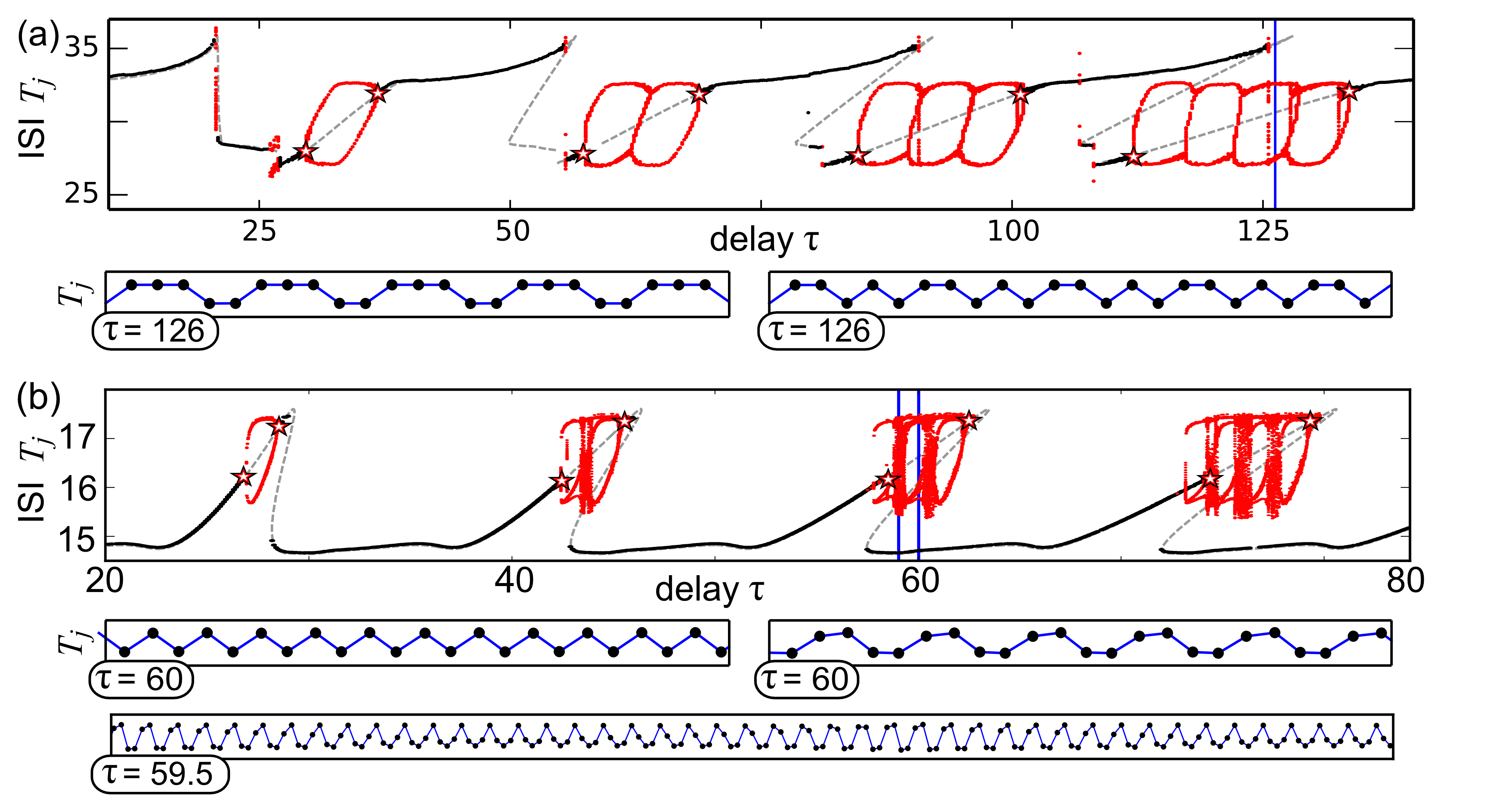}

\protect\caption{\label{fig:bifdiags-HH-FHN}Bifurcation diagrams for
(a) an electrically implemented FitzHugh-Nagumo system and (b) a simulated
Hodgkin-Huxley neuron model. RS solutions are indicated by black dots,
solutions with more than one ISI by red dots. The branches of RS solutions
obtained from the measured PRCs and (\ref{eq:T_RS}) are plotted as
dashed grey lines. Multi-jitter bifurcations are indicated by star-shaped
markers. The values of the delay $\tau$ are given in ms. The panels
below the diagrams show examples of stable bipartite solutions (ISIs
versus time). The values of $\tau$ for which the solutions exist
are also indicated by vertical blue lines in the corresponding diagrams.}
\end{figure}
In order to evaluate the practical relevance of the theory described
above we consider two realistic systems: (i) an electronic implementation
of the FitzHugh-Nagumo oscillator \cite{FitzHugh1961,Binczak2003,shchapin2009dynamics,Klinshov2014}
with time-delayed pulsatile feedback, and (ii) a numerically simulated
Hodgkin-Huxley model \cite{Hodgkin1952} with a delayed, inhibitory,
chemical synapse projecting onto itself \cite{Foss2000,Hashemi2012}.
 A detailed description of the systems is given in the Appendix {[}Secs.~\ref{sec:s2} and \ref{sec:s3}{]}. 
 In both cases the measured PRC exhibits parts with slope less than $-1$ {[} see Figs.~\ref{fig:s5} 
 and \ref{fig:s6}{]}. Therefore the
existence of multi-stable jittering can be conjectured on the basis
of our results for system (\ref{eq:1}). Figure~\ref{fig:bifdiags-HH-FHN}
presents experimentally (for the FitzHugh-Nagumo oscillator) and numerically 
(for the Hodgkin-Huxley model) obtained bifurcation diagrams
showing ISIs for varying delays.  Both systems clearly show that
a stable RS solution destabilizes closely to the multi-jitter bifurcation
points. Where the RS regime is unstable, the system switches to irregular
spiking, and we mainly observe $(P+1)$-periodic bipartite solutions.
The insets in the lower part of Fig.~\ref{fig:bifdiags-HH-FHN}(a)
show two such period-5 bipartite regimes of the FitzHugh-Nagumo oscillator.
Note that both of them coexist at $\tau=126$ ms, which illustrates
the multistability of the system. Similarly, two period-4 bipartite
regimes coexisting for $\tau=60$ ms for the Hodgkin-Huxley model are shown in \ref{fig:bifdiags-HH-FHN}(b).

Bipartite solutions are a basic form of jittering both in phase-reduced
models and realistic systems. However, beyond the multi-jitter bifurcation
the bipartite solutions may undergo subsequent bifurcations. In system
(\ref{eq:1}), we observed higher-periodic, quasiperiodic and chaotic
regimes for larger steepnesses of the PRC ($q>70$). Similar regimes
were also found for the Hodgkin-Huxley model. An example showing aperiodic
jittering is shown in Fig. ~\ref{fig:bifdiags-HH-FHN}(b) for $\tau=59.5$
ms. In the Appendix we show more examples of aperiodic
jittering {[}see Fig.~\ref{fig:s7}{]}.

To conclude, in a phase oscillator with delayed pulsatile feedback
(\ref{eq:1}) we discovered a surprising bifurcation leading to the
emergence of a large number ($\sim\exp\tau$) of jittering solutions.
We showed that this multi-jitter bifurcation does not only appear
in phase-reduced models, but also in realistic neuron models and even
in physically implemented electronic systems. These findings support
our theoretical results and provide motivation for a deeper study
of the multi-jitter phenomenon.

The possibility of jittering depends on the steepness of the PRC which
is an easily measurable quantity for most oscillatory systems \cite{Winfree2001}.
Thus, our findings provide an easy criterion to check for the existence
of jittering in a given system. This may prove useful in a variety
of research areas, where pulsatile feedback or interactions of oscillating
elements takes place. For instance, this might be one of the mechanisms
behind the appearance of irregular spiking in neuronal models with
delayed feedback \cite{Ma2007} and timing jitter in semiconductor
laser systems with delayed feedback \cite{Otto2012}. For applications
which exploit complex transient behavior such as liquid state machines
\cite{LSM} the high dimension of the unstable manifold at the bifurcation
can be interesting. Furthermore, in view of the possibility of a huge
number of coexisting attracting orbits beyond the bifurcation the
system can serve as a memory device by associating inputs with the
attractors to which they make the system converge.

\begin{acknowledgements}
The theoretical study was supported by the Russian Foundation for Basic Research (Grants No. 14-02-00042 and No. 14-02-31873), the German Research Foundation (DFG) in the framework of the Collaborative Research Center SFB 910, the German Academic Exchange Service (DAAD, research fellowship A1400372), and the European Research Council (ERC-2010-AdG 267802, Analysis of Multiscale Systems Driven by Functionals). The experimental study was carried out with the financial support of the Russian Science Foundation (Project No. 14-12-01358). 
\end{acknowledgements}

\newpage{}

\newpage{}

\begin{appendix}
\renewcommand{\theequation}{\Alph{section}.\arabic{equation}}
\renewcommand{\thefigure}{\Alph{section}.\arabic{figure}}
 \setcounter{equation}{0}
 \setcounter{figure}{0}

\section{Maps and basins for $P=1,2,3,4$}\label{sec:s1}

In this section we illustrate the ISI maps for cases $P=1,2,3,4$.
In all cases we use $Z_{{\rm ex}}(\varphi):=0.1\sin^{q}\left(\pi\varphi\right)$
with $q=28$.

$\boldsymbol{P=1.}$\textbf{ }In this case the map is one-dimensional:
\begin{equation}
T_{j+1}=1-Z\left(\tau-T_{j}\right).\label{eq:map-P1}
\end{equation}

\begin{figure}[h]
\begin{centering}
\includegraphics{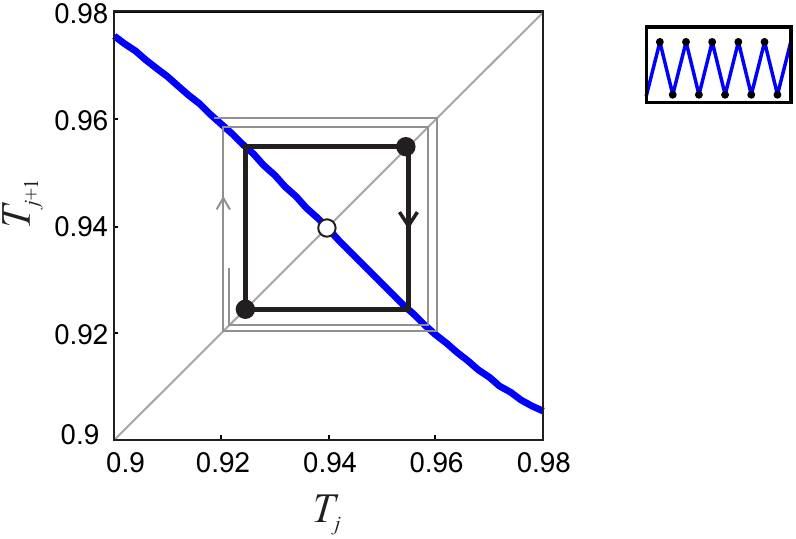}
\par\end{centering}

\protect\caption{\label{fig:s1}Cob-web diagram of (\ref{eq:map-P1}) for $\tau=1.5$.
The black solid dots correspond to a stable period 2 solution and
the white hallow dot corresponds to unstable regular spiking.}
\end{figure}

The dynamics of the map can be illustrated on a coweb diagram which
is depicted in Fig.~\ref{fig:s1} for $\tau=1.5.$ For this value of the delay,
the only attractor is a stable period 2 solution {[}cf. Fig.~2 of
the main text{]}.

\newpage{}

$\boldsymbol{P=2.}$ In this case the map is two-dimensional and reads

\begin{equation}
T_{j+1}=1-Z\left(\tau-T_{j}-T_{j-1}\right).\label{eq:map-P2}
\end{equation}
This can be equally rewritten in the vector form:

\[
\left(\begin{array}{c}
T_{1}\\
T_{2}
\end{array}\right)\mapsto\left(\begin{array}{c}
T_{2}\\
1-Z\left(\tau-T_{2}-T_{1}\right)
\end{array}\right).
\]

\begin{figure}[h]
\begin{centering}
\includegraphics{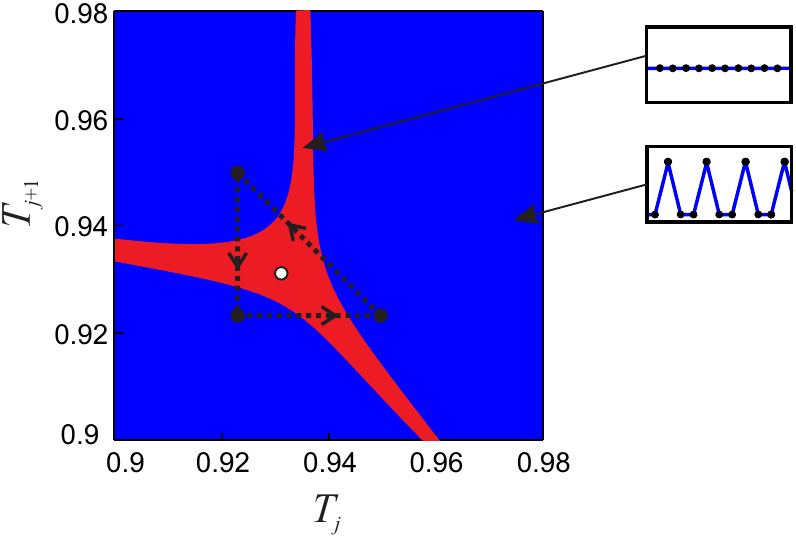}
\par\end{centering}

\protect\caption{\label{fig:s2}Phase plane of (\ref{eq:map-P2}) for $\tau=2.414$.
The black solid dots correspond to a stable period 3 solution and
the white hallow dot corresponds to stable regular spiking. The blue
region is the attraction basin of the period-3 solutions, red of regular
spiking.}
\end{figure}

Figure~\ref{fig:s2} shows attractors and their attraction basins of the map
(\ref{eq:map-P2}) for $\tau=2.414$. For this value of the delay
we observe a coexistence of a stable regular spiking solution and
a stable irregular period-3 solution of the form $(\overline{\Theta_{1},\Theta_{1},\Theta_{2}})$
{[}cf. Eq.~\ref{eq:T1_T2} of the main text{]}. 

\newpage{}

\begin{figure}[h]
\begin{centering}
\includegraphics{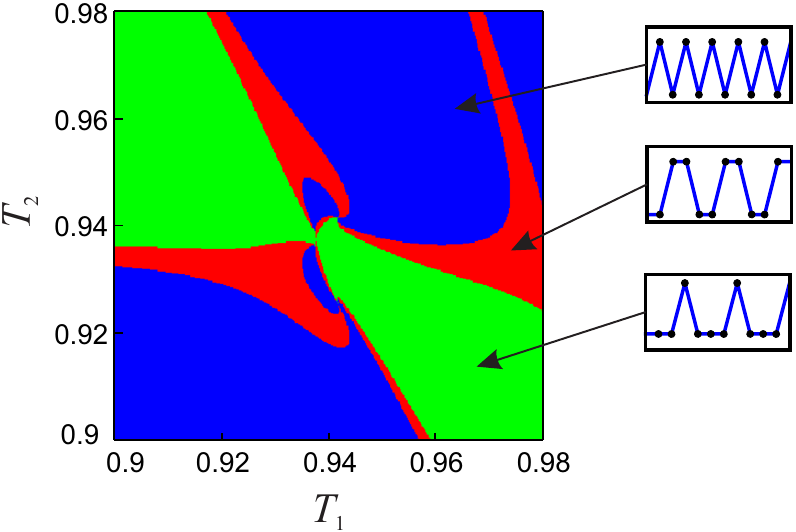}
\par\end{centering}

\protect\caption{\label{fig:s3}Planar section of the attractor basins of the
map (\ref{eq:map-P3}) by the plane (\ref{eq:plane1}) for $\tau=3.37$.
The corresponding attracting bipartite solutions are depicted in the
right part.}
\end{figure}

$\boldsymbol{P=3.}$ In this case the map is three-dimensional and
reads

\begin{equation}
T_{j+1}=1-Z\left(\tau-T_{j}-T_{j-1}-T_{j-2}\right),\label{eq:map-P3}
\end{equation}
or, written in an equivalent vector form
\[
\left(\begin{array}{c}
T_{1}\\
T_{2}\\
T_{3}
\end{array}\right)\to\left(\begin{array}{c}
T_{2}\\
T_{3}\\
1-Z\left(\tau-T_{3}-T_{2}-T_{1}\right)
\end{array}\right).
\]
For $\tau=3.37$ three different stable bipartite solutions of this
map coexist: $(\overline{\Theta_{1},\Theta_{1},\Theta_{1},\Theta_{2}})$,
$(\overline{\Theta_{1},\Theta_{1},\Theta_{2},\Theta_{2}})$, and $(\overline{\Theta_{1},\Theta_{2}})$.
Figure~\ref{fig:s3} depicts the basins of attractions confined to the two-dimensional
plane 
\begin{equation}
H=\{(T_{1},T_{2},T_{3})\,|\,\,T_{1},T_{2}\in[0.9,0.98],T_{3}=T_{1}\}\label{eq:plane1}
\end{equation}
of the three-dimensional phase space.

\newpage{}

\begin{figure}[h]
\begin{centering}
\includegraphics[bb=0bp 0bp 229bp 168bp,clip]{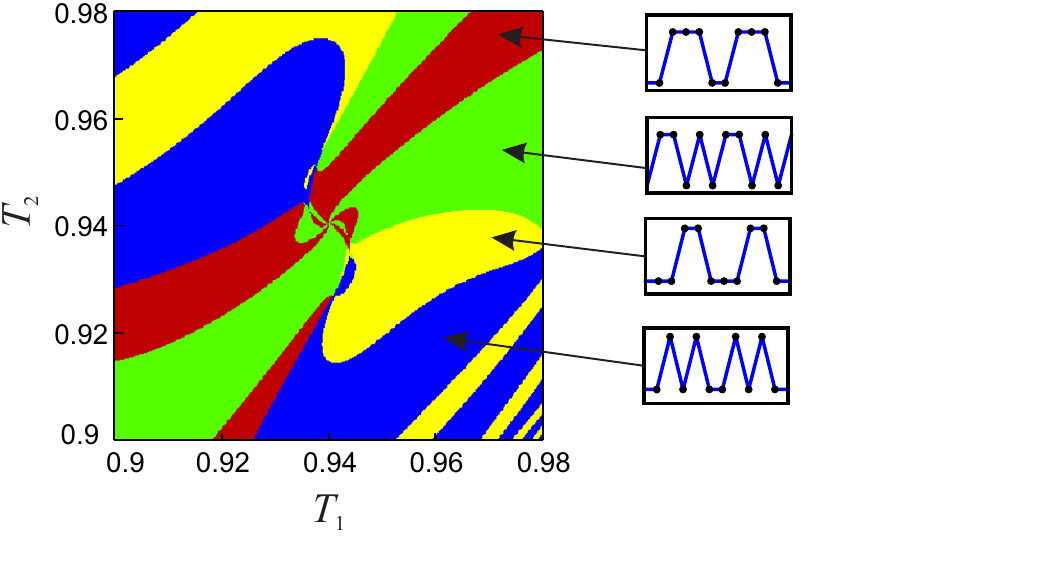}
\par\end{centering}

\protect\caption{\label{fig:s4}Planar section of the attractor basins by the
plane (\ref{eq:plane2}) of the map (\ref{eq:map-P4}) for and $\tau=4.32$.
The corresponding attracting bipartite solutions are depicted in the
right part.}
\end{figure}

$\boldsymbol{P=4.}$ In this case the map is four-dimensional and
has the form

\begin{equation}
T_{j+1}=1-Z\left(\tau-T_{j}-T_{j-1}-T_{j-2}-T_{j-3}\right).\label{eq:map-P4}
\end{equation}
Here we omit the vector form for brevity. For $\tau=4.32$ four different
stable bipartite solutions coexist: $(\overline{\Theta_{1},\Theta_{1},\Theta_{2},\Theta_{2},\Theta_{2}})$,
$(\overline{\Theta_{1},\Theta_{2},\Theta_{1},\Theta_{2},\Theta_{2}})$,
$(\overline{\Theta_{1},\Theta_{1},\Theta_{1},\Theta_{2},\Theta_{2}})$,
and $(\overline{\Theta_{1},\Theta_{1},\Theta_{2},\Theta_{1},\Theta_{2}})$.
An intersection of the attractors basins with the 2-dimensional plane
\begin{equation}
H=\{(T_{1},T_{2},T_{3},T_{4})|\,T_{1},T_{2}\in[0.9,0.98],T_{3}=T_{1},T_{4}=T_{2}\}\label{eq:plane2}
\end{equation}
 is shown in Fig.~\ref{fig:s4}.

\newpage{}

\section{Electronic FitzHugh-Nagumo oscillator}\label{sec:s2}
 \setcounter{equation}{0}
 \setcounter{figure}{0}

\begin{figure}[h]
\begin{centering}
\includegraphics[width=0.75\columnwidth]{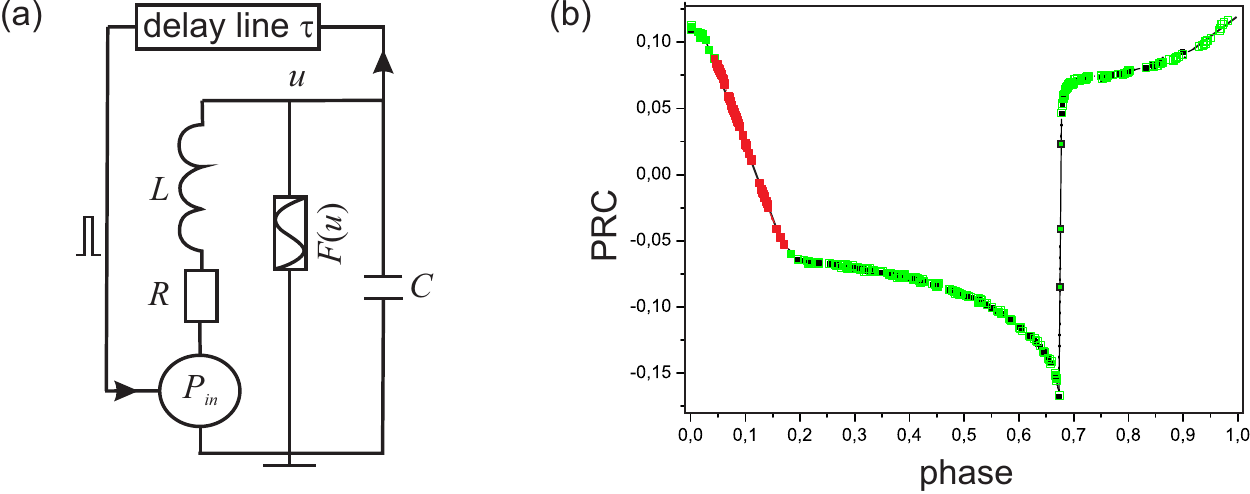}
\par\end{centering}

\protect\caption{\label{fig:s5}(a) Circuitry of the Fitzhugh-Nagumo oscillator.
(b) The measured PRC of the FitzHugh-Nagumo oscillator. The interval
with the slope less than minus one is highlighted in red.}
\end{figure}

The circuitry of the electronic FitzHugh-Nagumo oscillator as used
in the experiment {[}cf. Fig.~\ref{fig:bifdiags-HH-FHN}(a) of the main text{]} is depicted
in Fig.~\ref{fig:s5}(a), see Refs.~\cite{shchapin2009dynamics, Klinshov2014} for details. 
Here, $R=1$k$\Omega$,
$C=47$nF, $L=103.4$H, $P_{in}$ is an input from the delay line,
and $F(u)=\alpha u(u-u_{0})(u+u_{0})$ is the current-voltage characteristic
of the nonlinear resistor with $\alpha=2.02\times10^{-4}\Omega^{-1}\mbox{V}^{-1}$
and $u_{0}=0.82$V. The autonomous oscillations have period $T\approx30$ms.
The delay line is realized as a chain of monostable multivibrators.
A pulse of the amplitude $A=5$V and duration $\theta=0.42$ms is
delivered with a delay $\tau$ each time the voltage $u$ reaches
the threshold $u_{th}=-0.7$V. 

The noise level in the circuit was $\approx-40$dB (this means fluctuations
of $\approx20$mV for a signal amplitude of $\approx2$V).

The measured phase resetting curve for the given parameters is depicted
in Fig.~\ref{fig:s5}(b). The interval where the PRC slope is less than minus
one is highlighted in red. 

\newpage{}

\section{Hodgkin-Huxley model}\label{sec:s3}
 \setcounter{equation}{0}
 \setcounter{figure}{0}

The periodically spiking Hodgkin-Huxley neuron model, which was used
for the numericalresults in Fig.~\ref{fig:bifdiags-HH-FHN}(b) of the main text, is given
by the following set of equations \cite{Foss2000, Hashemi2012, Hodgkin1952} 

\begin{align}
C\dot{V}(t) & =I-g_{Na}m^{3}h(V(t)-V_{Na})-g_{K}n(V(t)-V_{K})\nonumber \\
 & -g_{l}(V(t)-V_{l})-\kappa(V(t)-V_{r})s(t-\tau),\label{eq:HH-voltage-eq}\\
\dot{m}(t) & =\alpha_{m}(V(t))(1-m(t))-\beta_{m}(V(t))m(t),\nonumber \\
\dot{h}(t) & =\alpha_{h}(V(t))(1-h(t))-\beta_{h}(V(t))h(t),\nonumber \\
\dot{n}(t) & =\alpha_{n}(V(t))(1-n(t))-\beta_{n}(V(t))n(t),\nonumber \\
\dot{s}(t) & =5(1-s(t))/(1+\exp(-V(t)))-s(t),\nonumber 
\end{align}
where $V(t)$ models the membrane potential, $\alpha_{m}(V)=(0.1V+4)/(1-\exp(-0.1V-4))$,
$\beta_{m}(V)=4\exp((-V-65)/18)$, $\alpha_{h}=0.07\exp((-V-65)/20)$,
$\beta_{h}(V)=1/(1+\exp(-0.1V-3.5))$, $\alpha_{n}(V)=(0.01V+0.55)/(1-\exp(-0.1V-5.5))$,
$\beta_{n}(V)=0.125\exp((-V-65)/80)$, $C=1\mu\text{F}/\text{cm}^{2}$,
$I=10\mu\text{A}/\text{cm}^{2}$, $g_{Na}=120\text{mS}/\text{cm}^{2}$,
$V_{Na}=50\text{mV}$, $g_{K}=36\text{mS}/\text{cm}^{2}$, $V_{K}=-77\text{mV}$,
$g_{l}=0.3\text{mS}/\text{cm}^{2}$, $V_{l}=-54.5\text{mV}$, $V_{r}=-65\text{mV}$,
and $\kappa=0.38\text{mS}/\text{cm}^{2}$.

\begin{figure}[h]
\begin{centering}
\includegraphics[width=0.4\columnwidth]{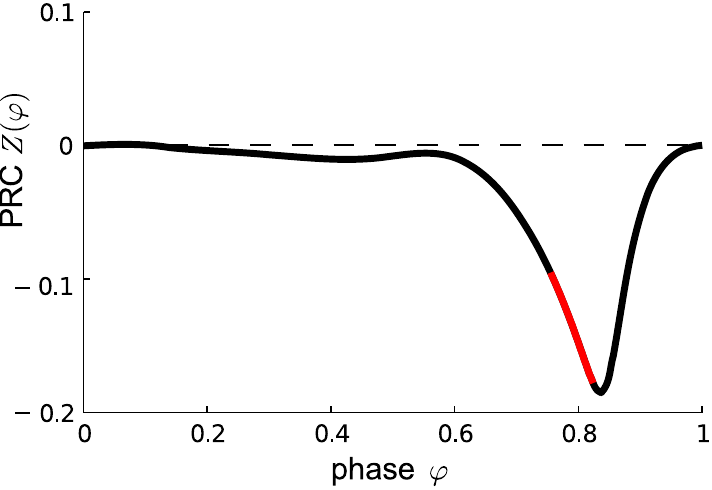}
\par\end{centering}

\protect\caption{\label{fig:s6}PRC of the Hodgkin-Huxley model (\ref{eq:HH-voltage-eq}).
The region, where $Z^{\prime}\left(\varphi\right)<-1$ is indicated
by red.}
\end{figure}

Figure~\ref{fig:s6} shows the PRC of (\ref{eq:HH-voltage-eq}), which was
measured by replacing the delayed feedback by an external stimulation
line through which singular synaptic pulses sampled from the same
system were applied at different phases.

\section{Emergence of chaotic jittering in the Hodgkin-Huxley model}\label{sec:s4}
 \setcounter{equation}{0}
 \setcounter{figure}{0}

\begin{figure}[h]
\includegraphics[width=1\textwidth]{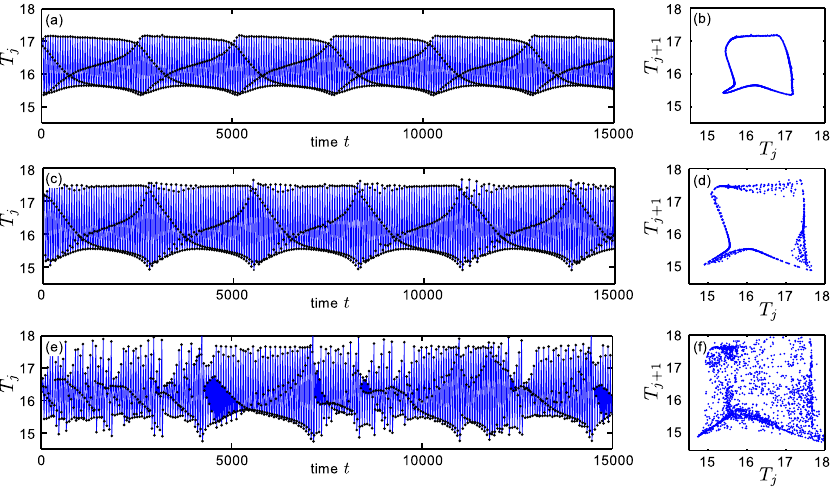}

\protect\caption{\label{fig:s7}Emergence of chaotic jittering in (\ref{eq:HH-voltage-eq}).
The plots (a), (c), and (e) show ISI sequences of trajectories observed
at the indicated feedback strengths $\kappa=0.38,\,0.4,\,0.41$. (a),(b):
Quasiperiodic jittering for $c=0.38$; (c),(d): weakly chaotic jittering
for $c=0.4$; (e),(f): clearly recognizable chaotic jittering for
$c=0.41$.}
\end{figure}

\begin{figure}[h]
\begin{centering}
\includegraphics[width=1\textwidth]{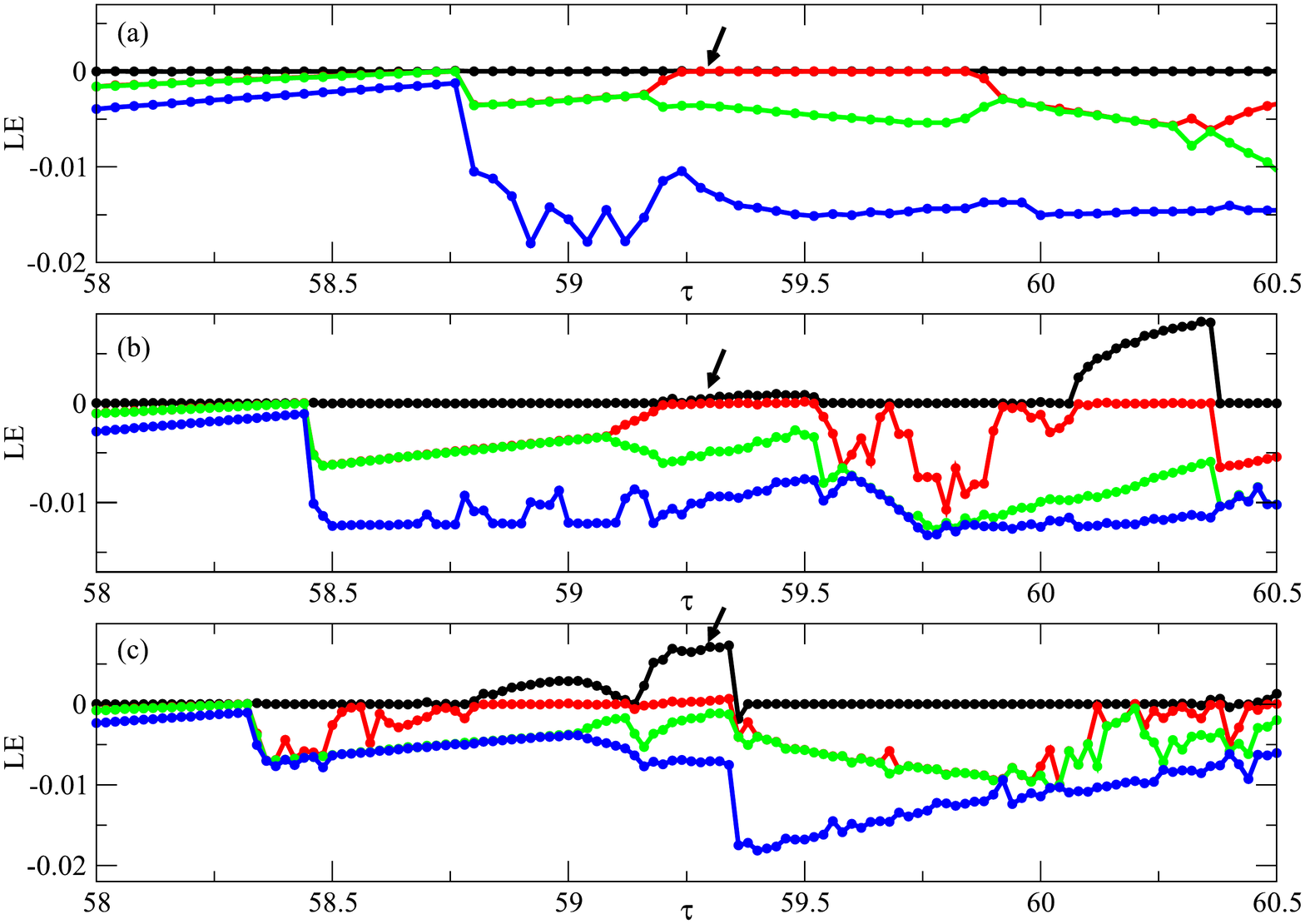}
\par\end{centering}

\protect\caption{\label{fig:s8}Four largest Lyapunov exponents of (\ref{eq:HH-voltage-eq})
with $\tau\in[58,60.5]$ and feedback strengths (a) $\kappa=0.38$;
(b) $\kappa=0.4$; (c) $\kappa=0.41$. Arrows indicate the delays,
for which the trajectories in Fig.~\ref{fig:s7} are shown.}
\end{figure}

Figure~\ref{fig:s7} illustrates the emergence of chaotic jittering states for
increasing feedback strength $\kappa$. It shows three different trajectories
for $\tau=59.3$ and $\kappa=0.38,\,0.4,\,0.41$. In plot (a), the
ISIs of a quasiperiodic solution are shown for $c=0.38$. A black
dot is placed at $(t_{j},T_{j})$, where $t_{j}$ is the moment when
the $j$-th ISI ends and $T_{j}$ is its duration. Subsequent dots
are joined by a blue line. The sequence is contained in a torus in
the phase-space, whose projection to the $(T_{j},T_{j+1})$-plane
is shown in plot (b). For each $T_{j}$ from the sequence of approximately
thousand observed ISIs, a blue dot was placed at the coordinates $(T_{j},T_{j+1})$.
Plots (c) and (d) depict a solution close to the onset of chaotic
jittering for $c=0.4$. The corresponding Lyapunov exponent is positive
but small {[}see Fig.~\ref{fig:s8}(b){]}. A more pronounced chaos is exhibited
by the solution existing at $c=0.41$, which is shown in plots (e)
and (f). Note that the emergence of chaos is accompanied by a loss
of smoothness of the torus consisting of the quasiperiodic trajectories 
\cite{Afra2003}.

Figure~\ref{fig:s8} shows the numerically calculated Lyapunov exponents (LE)
for different values of the feedback strength $\kappa$. For each
value of $\kappa$ the four largest exponents are shown. Starting
from $\tau=58$ on the RS solution with maximal period {[}cf. Fig.~\ref{fig:bifdiags-HH-FHN}(b)
for $\kappa=0.38${]} each computation for $\tau\in[58,60.5]$ was
initialized with a solution on the previous attractor as initial data.
Note that there exists always one LE which has real part zero, since
the corresponding attractors are not steady states. For each value
of $\tau$, the point where the multi-jitter bifurcation occurs can
clearly be recognized as all four depicted LE approach zero nearly
at the same value of $\tau=\tau_{A}$ ($\tau_{A}\approx58.75$ for
$\kappa=0.38$, $\tau_{A}\approx58.4$ for $\kappa=0.4$, and $\tau_{A}\approx58.3$
for $\kappa=0.41$). Dynamics found in the depicted range beyond these
points, i.e. for $\tau_{A}<\tau<60.5$, are irregularly spiking regimes.
For $\kappa=0.38$, two LE are zero in the interval $\tau\in[59.25,59.8]$,
which indicates a torus, i.e. quasiperiodic behavior as illustrated
in Fig.~\ref{fig:s7}(a). For all other values of $\tau$ in the depicted range,
we observe periodic solutions which exhibit ISI sequences of period
4 {[}see Fig.\ref{fig:bifdiags-HH-FHN}(b) of the main text{]}. When the feedback strength
is increased to $\kappa=0.4$ the dynamics in the corresponding interval
$\tau\in[59.2,59.5]$ become weakly chaotic. For $\kappa=0.41$ the
LE become larger in the interval $\tau\in[58.75,59.2]$ which indicates
a more pronounced chaos. Note that even for feedback strengths where
chaotic jittering is observed, periodic solutions still exist at other
values of $\tau$.

\end{appendix}

\end{document}